# DevOps in Practice – A preliminary Analysis of two Multinational Companies

Jessica Díaz[1][0000-0001-6738-9370] Jorge E. Perez[1][0000-0003-3349-6017] Agustín Yague[1][0000-0002-4761-0901] Andrea Villegas[1,2] and Antonio de Antona[3]

[1] Universidad Politécnica de Madrid, ETSI de Sistemas Informáticos, 28031 Madrid, Spain
[2] Sistemas Avanzados de Tecnología, S. A. (SATEC), 28023 Madrid, Spain
[3] Everis Spain S.L., Madrid, Spain
(yesica.diaz, jorgeenrique.perez,agustin.yague)@upm.es,
a.villegas@alumnos.upm.es

**Abstract.** DevOps is a cultural movement that aims the collaboration of all the stakeholders involved in the development, deployment and operation of software to deliver a quality product or service in the shortest possible time. DevOps is relatively recent, and companies have developed their DevOps practices largely from scratch. Our research aims to conduct an analysis on practicing DevOps in +20 software-intensive companies to provide patterns of DevOps practices and identify their benefits and barriers. This paper presents the preliminary analysis of an exploratory case study based on the interviews to relevant stakeholders of two (multinational) companies. The results show the benefits (software delivery performance) and barriers that these companies are dealing with, as well as DevOps team topology they approached during their DevOps transformation. This study aims to help practitioners and researchers to better understand DevOps transformations and the contexts where the practices worked. This, hopefully, will contribute to strengthening the evidence regarding DevOps and supporting practitioners in making better informed decisions about the return of investment when adopting DevOps.

## 1 Introduction

In the recent digital history, it is possible to verify that success is not always achieved by the product that is better built, more usable, or of better quality, but by the one that appears first and meets a certain need. This is why the software industry tries to be more agile, more tolerant to change, more adaptable to new needs, and above all, tries to shorten development time from request to implementation. Companies that can release software early and frequently have a higher capability to innovate and compete in the market. Innovative companies, such as Google, Amazon, Netflix, LinkedIn, Facebook, and Spotify, initiated an organizational transformation that aimed fast speed in releases and quick response time to customer demands. DevOps is an organizational transformation that had its origin at the 2008 Agile conference in Toronto, where P. Debois highlighted the need of resolving the conflict between developer teams and operation teams when they have to collaborate to provide this quick response time to customer demands [1]. Later, at the O'Reilly Velocity Conference, two Flickr employees delivered a seminal talk known as "*10+ Deploys per Day: Dev and Ops Cooperation at Flickr*" which can be considered the starting point to extend agile beyond development [2]. Today an entire industry has been created around DevOps tools whose objective is to automate best practices, such as continuous delivery and continuous deployment, that promote fast and frequent delivery of new and changing features while ensuring quality and non-disruption of the production environment and customers [3].

But beyond all that, DevOps is a cultural movement that aims the collaboration among all the stakeholders involved in the development, deployment and operation of software to deliver a high-quality product or service in the shortest possible time. DevOps breaks down organizational silos and "*stresses empathy and cross-functional collaboration within and between teams—especially development and IT operations—in order to operate resilient systems and accelerate delivery of changes*" [4]. It is a simple concept, but its adoption by organizations is enormously complicated because of great differences in the way in which DevOps promotes to work and the traditional way in which most software companies have been working for decades. DevOps is founded on the Lean principles and shares its values, such as process optimization, search for continuous improvement, and the enhancement of customer satisfaction.

However, DevOps is relatively recent, and little is known about best practices and the real value and barriers associated with DevOps in industry. Companies have developed their DevOps practices largely from scratch—by training employees on the fly. Google, Amazon, Netflix, LinkedIn, Facebook, and Spotify are some examples of successful companies whose DevOps practices have been reported and disclosed in IT books, blogs and events. These provide a valuable information for companies, however, for most of them, it is quite difficult to match these leader companies and adopt the practices they disclose. In addition, failures are not described. Some annually reports about the state of DevOps, such as the report made by DORA (DevOps Research & Assessment association) [5] and the report made by Puppet and Splunk [6], analyze data of survey questionnaires over 30,000 technical professional worldwide, respectively. The first one identifies a set of *software delivery performance profiles* (elite, high, medium and low performance) and relates DevOps practices with these profiles. The second one identifies 5 stages of DevOps evolution (aka. the *DevOps evolutionary model*) and establishes the practices that define and/or contribute to success in that stage. These reports also provide a valuable information for companies as they provide a global picture; however, the wide range of participant companies and the great variability among participants make difficult, for a company, to find the right way for a DevOps transformation based on similarities (e.g. IT department size, business, scope, DevOps Teams size, DevOps strategy, etc.) with other companies. Erich et al. [7] and Lwakatare et al. [8] also performed exploratory studies on six companies and one company, respectively, providing a key baseline for future studies with a broader scope until achieving the saturation for qualitative studies.

Our research aims to conduct an analysis on practicing DevOps in +20 software-intensive companies to provide patterns of DevOps practices and identify their benefits and barriers. This paper presents the preliminary analysis of an exploratory case study based on the interviews to relevant stakeholders of two (multinational) companies. DevOps embodies a vast and diverse set of practices, from which some patterns can be generalized under certain conditions, depending on the environment [9]. The analysis of these two case studies may help researchers and professionals to understand the barriers and benefits (specifically, delivery software performance) when two companies of the software industry made a DevOps transformation, how these companies dealt with the transformation (specifically, DevOps team topology), and finally, it may help others to make better informed decisions based on this know-how. There are some decisions that can lead to the failure of an organization, and many others to success, so that the only way to be sure of being on the right way is to follow one that has been successfully proven on numerous occasions.

## 2       Exploratory Case Study

The research methodology has been previously described, discussed, and improved at the *Fostering More Industry-Academic Research in XP* (FIAREX) workshop, part of XP 2018 conference [10]. We have followed the guidelines for conducting case study research in software engineering proposed by Runeson and Höst [11]. We have established a chain of evidence by following a strict process that consists of the preparation of a questionnaire and interviews, performance and recording, transcription, coding, and analysis. To qualitatively analyze the data, we have used the thematic analysis approach [12][13], which is one of the most used synthesis methods that consists of coding, grouping, interconnecting and obtaining patterns. The last two activities were also supported using the clustering technique, which divides samples in groups called clusters based on their similarity. The visualization of these clusters helped us to better interpret and relate the qualitative data.

### 2.1     Data Collection and Instruments

The interviews were conducted face-to-face by two researchers. The interview consists of 100 questions and takes about 2.5 hours. The questions were collected from the existing literature conducting survey studies on DevOps state [5][6][14], exploratory studies [7][8], as well as from meetings with experts in some international and national workshops (e.g. at the FIAREX workshop part of the XP conference [10] and a local industrial workshop organized by the authors[1]) and national events (e.g. DevOps Spain[2] and itSMF events[3]). The interview is structured to collect professional 4information from interviewees, organizations, DevOps adoption processes, DevOps teams' topology, culture related practices, team related practices, collaboration related practices, sharing related practices, automation related practices, measurement and monitoring related practices, barriers, and results. This questionnaire includes a set of short, open, and semi-open questions in which the interviewee can choose one or more options, explain their selections, or add a new answer. Both options and questions have been refined as we gained more knowledge during the interviews, the workshops and the events. An example is question 17 about the DevOps teams' topology and its scope within the IT department. It was initially an open question, but after 4-5 interviews we realized that answers were too long and not clarified the topology, so we added some options based on the *DevOps Topologies collection of patterns* by Matthew Skelton and Manuel Pai [15] and the *organizational structures used in DevOps journey* by the State of DevOps Report [6]. After analyzing more interviews, we defined our own DevOps Teams patterns (see Section 2.2).

---

[1] (Spanish) Workshop on DevOps located at Universidad Politécnica de Madrid, Spain, https://www.youtube.com/watch?time_continue=6&v=rDHv3dK_Am8, last accessed 2019/08/01

[2] https://www.devops-spain.com/  last accessed 2019/08/01

The interview also asks for the *deployment frequency*, i.e., the number of deploys to production of an application per unit of time; the *lead time for changes*, i.e., the time from a change in the code to code is successfully running in production; and the *time to recovery*, i.e., elapsed time to restore a service when an incident causes its unavailability. These metrics were defined by DORA as indicators for defining a set of *software delivery performance profiles* (elite, high, medium and low performance) [5]. We have adapted the scale that is used for these indicators to classify companies according to the profiles by DORA for the scale that is shown is Table 1. This work is required because, as mentioned before, this kind of reports analyzes massive data, and the variability of these data is huge (e.g. the lead time goes from less than one hour to six months, and the data we have managed for lead time none exceed one day). Additionally, we limited the lead time for changes affecting to one line of code as we think that asking for the lead time of a change in the code is ambiguous.

**Table 1.** Software Delivery Performance indicators.

| Software delivery performance indicators | Elite | High | Medium | Low |
| --- | --- | --- | --- | --- |
| Deployment frequency | On demand, multiple deploys per day | One deploy per day | One deploy per week | Between once per week and one per month |
| Lead time for changes | Less than one hour | Less than one hour | Between one hour and one day | Between one hour and one day |
| Mean Time to Recovery | Less than one hour | Less than one hour | Between one hour and one day | Between one day and one week |

### 2.2 Subjects

This paper focuses on two companies (ID17 and ID18). In these interviews three people were interviewed: a consultant from Everis with +6 years of experience and +4

---

[3] https://www.itsmf.es/index.php?option=com_content&view=article&id=3133:2018-10-11-00-30-06&catid=79:noticias&Itemid=401, last accessed 2019/08/01

years in DevOps that worked for ID17 and ID18, the director of the DevOps department from organization ID17 with +12 years of experience and +6 years in DevOps, and a Scrum master from organization ID18 with +15 years of experience and +2 years in DevOps. Table 2 shows the description of these organizations. ID17 is a large company whose structure is very departmental: DevOps department (22 people), operation & cloud systems (12), operation and on-premise systems (15-20), security (12), architecture (20), quality assurance (10), service/help desk (22), a number of development departments (4-50 people) composed by *squads* (4-9 people). Squads are similar to Scrum teams, i.e. are the basic unit of development at Spotify, who coined this concept[4]. It is necessary to highlight that these teams have also the appropriate skills to release to production. This company also adopted the concept of *chapter* to designate people having similar skills and working within the same general competency area in different squads. This company has a DevOps chapter, and Architecture chapter, and QA chapter. ID 18, despite its small size, also has departmental structure, with different departments for development, DevOps & Cloud, QA, and security.

**Table 2.** Organizations' subject description.

| ID | Scope | Organization Size | Business | Creation year | IT department size | DevOps Team Num & Size |
|---|---|---|---|---|---|---|
| 17 | International | Large[5] | Telecommunications | Between 2000 and 2010 | 500 | 1 Team (22 members) |
| 18 | International | Large | Real state | Before 2000 | 30 | 1 Team (5 members) |

## 3  Key Findings

**RQ1** *What problems do companies try to solve and what results try to achieve by implementing DevOps*? ID 17 disclosed that the organization size, the diversity of its departments (development, operations, security, service, QA, architecture, etc.) as well as the interaction between them, and the complexity of its processes, hampered reducing time to market, and made this company less competitive. ID18 disclosed that the organization devoted most of the time to maintaining legacy applications and when this organization decided substitute the core legacy application with a new one, the CEO decided to make a significant change in the methodology, interaction between teams and the delivery and releasing processes to reduce time to market.

**RQ2** *What are the DevOps practices according to software practitioners*? This paper focuses on team related practices. Based on data collected from this study (+20 organ-izations), we have defined four patterns that describe the topology of DevOps teams and their scope within the IT department (see Fig. 1):

(i) <u>*Interdepartmental DevOps teams' pattern*</u> represents a close collaboration between Dev teams and Ops teams although these teams belong to different departments with different managers. Other authors called this pattern as *Dev and Ops Collaboration* [15] although we have identified two modalities: a combination of DevOps and traditional teams/approaches (a bimodal approach for different product/services) and only DevOps teams but maintaining the departmental structure.

(ii) <u>*Native DevOps teams pattern*</u> represents a close and efficient collaboration between Dev teams and Ops teams (also QA, security, etc.). It is an approach mainly adopted by start-up companies in which there is not separated roles for dev and ops. Other authors called this pattern as *Fully Shared Ops Responsibilities* [15].

(iii) <u>*DevOps as a Service*</u> is typical for companies without enough staff or experience, or very departmental and large companies, which cannot initially afford a complete DevOps transformation. This pattern provides an especial DevOps chapter that facilitates and helps to spread awareness of DevOps practices. According to other topologies this pattern is considered an antipattern *DevOps Team Silo* that only has sense when the team is not permanent, lasting less than (say) 12 or 18 months [15]. If silos are broken, this pattern could be considered as *DevOps Advocacy Team* [15]. According to our study, the DevOps service team usually becomes a department with its own manager, however we did not observe the creation of a new silo. Additionally, in our study no organization outsourced this service (*DevOps as an External Service*).

(iv) <u>*Ops as a Service*</u> represents those situations in which the traditional IT Operations department assumes the DevOps competences mainly by automating infrastructure provision (and possibly other more processes) on which applications are deployed and run. According to other topologies, this pattern could be *Ops as Infrastructure-as-a-Service (Platform)* [15].

---

[4] https://blog.crisp.se/wp-content/uploads/2012/11/SpotifyScaling.pdf, last accessed 2019/08/01

[5] Spanish law 5/2015, on the promotion of business financing, states that a small company has a maximum of 49 workers and a turnover or total asset value of less than ten million euros; and medium-sized companies are those with less than 250 workers and a turnover of less than fifty million euros or an asset of less than 43 million euros. Meanwhile, large companies are those that exceed these parameters.

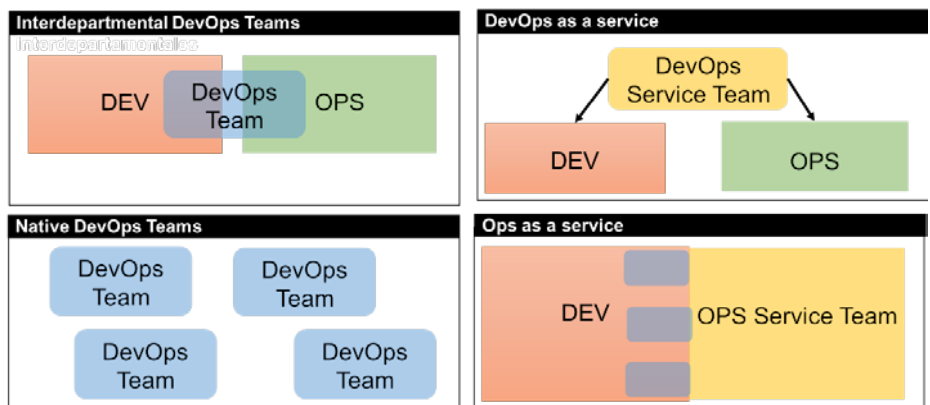

**Fig. 1.** DevOps team topology.

According to this classification, the organizations ID17 and ID18 implemented *DevOps as a Service* and *Ops as a service,* respectively. In the organization ID 17, the DevOps Service Team is a department composed by two squads. One of them automates processes and develops a DevOps platform for internal use, and the other one acts as a chapter so that, its members work closely with the development departments, evangelizing both DevOps practices and the use of its internal platform. In the organization ID 18, the traditional Ops (renamed as DevOps and Cloud department) provides services to other development departments. In this organization two models coexist: DevOps principles and practices for new developments and traditional approaches for core and legacy applications (bimodal approach).

**RQ3** *What were the achieved results of implementing DevOps*? Table 3 shows the data for the software delivery performance. According to these data and the benchmark of Table 1, we can say that organization ID17 has achieved a medium performance (medium deployment frequency, medium lead time, and low mean time to recovery) and ID18 also achieved a medium performance (low deployment frequency, medium lead time, and medium mean time to recovery).

**Table 3.** Results of software delivery performance indicators.

| ID | Deployment frequency | Lead Time for Changes | Mean Time to Recovery |
|---|---|---|---|
| 17 | One deploy per week | Between one hour and one day | Between one day and one week |
| 18 | One deploy per sprint (3 weeks) | Between one hour and one day | Between one hour and one day |

**RQ4** *What barriers are encountered when implementing DevOps*? ID17 disclosed about the misalignment among departments and the inflexibility of communication processes, whereas ID18 disclosed the complexity of standardizing and automating processes.

## 4 Conclusions and Threats to Validity

This paper presented the preliminary results of analyzing two organizations through an exploratory case study. The organizations were interviewed through a specific questionnaire to assess the state of DevOps. The data were systematically analyzed, and metrics were customized to have a better profiling of companies. The results mainly focused on analyzing the DevOps team topologies and the benefits when adopting DevOps in terms of software delivery performance. The defined questionnaire for interviews and the process defined to analyze these interviews provides a powerful tool to get results about the DevOps topics under research. The complete case study aims to tackle a significant number of software-intensive companies (+20) to give a detailed analysis of problems, barriers, benefits and practices patterns when organizations start a DevOps transformation, as well as of the relation between concepts (e.g. some practices and their resulting benefits). These patterns could provide a set of good practices when organizations decide to start DevOps transformation.

The main threat to validity is regarding with construct validity. Specifically, we used the *convenience sampling strategy*, which is a non-probability/non-random sam-pling technique used to create sample as per ease of access to organizations and the relevant stakeholders to the study. This could lead to organizations not fully reflecting the target audience.

## Acknowledgment

This work is supported by the project CROWDSAVING (TIN2016-79726-C2-1-R).